\documentstyle[tighten,floats,12pt,prd,aps]{revtex}

\newcommand{\beq}{\begin{equation}}
\newcommand{\eeq}{\end{equation}}

\def\lap{\lower.5ex\hbox{$\; \buildrel < \over \sim \;$}}
\def\gap{\lower.5ex\hbox{$\; \buildrel > \over \sim \;$}}

\input epsf
\begin{document}

\title{Gravity in the Randall-Sundrum Brane World.}
\author{Jaume Garriga$^{1,2}$ and Takahiro Tanaka$^{1,2,3}$\\$~$}
\address{$^{1}$IFAE, Departament de Fisica, Universitat Autonoma de Barcelona,\\
08193 Bellaterra $($Barcelona$)$, Spain$;$\\
$^{2}$Isaac Newton Institute, University of Cambridge\\
20 Clarkson Rd., Cambridge, CB3 0EH, UK$;$\\
$^{3}$Department of Earth and Space Science, Graduate School of Science\\
 Osaka University, Toyonaka 560-0043, Japan.
}

\maketitle

\thispagestyle{empty}
\vspace{2mm}
\begin{abstract}
We discuss the weak gravitational field created by isolated matter 
sources in the Randall-Sundrum brane-world. In the case of
two branes of opposite tension, 
linearized Brans-Dicke (BD) gravity is recovered on
either wall, with different BD parameters. On the 
wall with positive tension
the BD parameter is larger than 3000 provided that the separation between
walls is larger than 4 times the AdS radius. For the wall of negative 
tension the BD parameter is always negative. In either case,
shadow matter from the other wall gravitates upon us. For equal
Newtonian mass, light deflection from shadow matter is 25 \% weaker than
from ordinary matter. 
For the case of
a single wall of positive tension, Einstein gravity is recovered
on the wall to leading order, and if the source is stationary the field 
stays localized near the wall. We calculate the leading 
Kaluza-Klein corrections to the 
linearized gravitational field of a non-relativistic spherical object and 
find that the metric is different from the Schwarzschild solution at large
distances. We believe that our linearized solution corresponds to
the field far from the horizon after gravitational collapse of matter on
the brane.\\
\end{abstract}

\hspace*{1cm}
\newpage

It has recently been shown \cite{RS2} that something very similar to four 
dimensional Einstein gravity exists on a domain wall (or 3-brane) 
of positive tension which is 
embedded in a five dimensional anti-de Sitter space (AdS). 
The striking feature about this model is that an effective
dimensional reduction occurs without the need of compactifying 
the fifth dimension. The reason is that ``Kaluza-Klein'' (KK) excitations, 
which have nonvanishing momentum in the fifth 
direction, are suppressed near 
the brane. Thus, even though the
KK modes are light, they almost decouple from matter fields - which 
are constrained to live on the wall. Gravitational interactions amongst 
matter fields are mediated predominantly by the ``zero mode'', 
which is often described as a bound state of gravity on 
the wall. The case of two parallel domain walls, one with positive tension 
and another with negative tension, has also been discussed  
in an attempt to solve the much debated hierarchy problem \cite{RS1}.
The possibility that we may be living in a brane is rather 
tantalizing, and many questions
arise as to how gravity should look like in such a world. What are
the corrections to Einstein gravity? How does the ``shadow'' 
matter living in the other brane gravitate upon us? What is the final state of
gravitational collapse? Exploring some of these aspects will be the 
subject of the present paper.

Some attention has been devoted to cosmological \cite{cosm,SMS}, as well as
nonperturbative vacuum solutions \cite{CG,CHR} in this context. 
In Randall and Sundrum's solution the metric 
induced on the brane is flat. However, straightforward generalizations 
can be obtained in which the induced metric is any {\em vacuum} 
solution of the four dimensional Einstein's equations.
Generalizations of this sort where given in \cite{CG} and \cite{CHR}, 
where the plane wave and the Schwarzschild solutions were considered.
In these solutions,  the metric on every spacetime 
slice parallel to the brane is
the same as the metric on the brane, just rescaled
by the AdS conformal factor.
Thus, the gravitational field extends all the way to 
the AdS horizon, at infinite distance from the brane.
In the Schwarzschild case, the
five dimensional solution is a black string hidden behind a 
``cylindrical'' horizon extending to infinity. As shown 
by Chamblin et al. \cite{CHR}
tidal forces felt by freely falling observers actually 
become infinite as the AdS horizon is approached, which is not
very satisfactory from the physical point of view. However,
it was argued that since the infinite cylindrical horizon 
is unstable, the final state of collapse would perhaps
have a horizon in the shape of a cigar (rather than a full infinite cilinder).

Although some intuition can be drawn from the previous examples,
it would be interesting to find physical solutions where the
gravitational field stays localized near the sources.
For this 
purpose, an analysis of the weak gravitational field created by 
isolated matter sources on the brane seems to be the best starting point.
Let us begin with the case of a single membrane 
of positive tension embedded
in five dimensional AdS space. The metric is given by:
\begin{equation}
ds^2=g_{ab}dx^a dx^b=dy^2 + a^2(y)\eta_{\mu\nu}dx^\mu dx^\nu.
\label{metric}
\end{equation}
Here, $a(y)= e^{-|y|/\ell}$, where $\ell$ is the curvature radius of AdS,
and $\eta_{\mu\nu}$ is the Minkowski metric in four dimensions.
The cosmological
constant on the bulk is given by $\Lambda=-6 \ell^{-2}$ and
the wall tension is given by $\sigma = 3/ 4\pi \ell G_5$,
where $G_5$ is Newton's constant in five dimensions.

Denoting the perturbed metric by 
$
\tilde g_{ab}=g_{ab}+h_{ab},
$ 
the Randall-Sundrum (RS) gauge is defined by 
\begin{equation}
h_{55}=h_{\mu 5}=0, \quad h_{\mu}{}^{\nu}{}_{,\nu}=0, 
\quad h^{\mu}{}_{\mu}=0. 
\label{rsgauge}
\end{equation}
It is possible to show 
that these conditions can be chosen everywhere in the bulk\cite{effort}. 
In this gauge, the equations of motion take the simple form
\begin{equation}
\left[a^{-2}\Box^{(4)} + \partial_y^2 -4\ell^{-2}\right]
h_{\mu\nu}=0. \label{rseq}
\end{equation}
The advantage of this gauge is that all components of the metric 
are decoupled. 
However, in general, when we choose 
the gauge (\ref{rsgauge}) in the bulk,
the brane will not be located at $y=0$. Instead, as we shall see,
its location will be given by
$
y=-\hat\xi^5(x^\mu)~
$
(see fig.~1),
where $\hat\xi^5$ is the solution of the equation
\begin{equation}
\Box^{(4)} \hat\xi^5 = {\kappa \over 6} T.
\label{ywall}
\end{equation}
Here $T=T^{\mu}{}_{\mu}$, and $\kappa=8\pi G_5$. 
In our definition
of $T_{\mu\nu}$ we are not including the contribution from the wall itself.
To  proceed, it will be convenient to
go momentarily to Gaussian normal coordinates, 
which we denote by $\bar x^a$. 
By definition, the wall is located at $\bar y=0$, and 
we have $\bar h_{55}=\bar h_{\mu 5}=0$. 
Gaussian coordinates are also interesting for us
because $\bar h_{\mu\nu}(\bar y=0)$ is the metric perturbation 
induced on the wall. We impose even parity under $\bar y\to -\bar y$, 
and we shall work on the positive side 
in the following discussion. Then, 
the junction condition on the extrinsic 
curvature at the wall requires that
$
\partial_y (g_{\mu\nu}+ \bar h_{\mu\nu})
 =-(\kappa/ 3)[\sigma (\gamma_{\mu\nu}+\bar h_{\mu\nu})
  + 3 T_{\mu\nu}- T \gamma_{\mu\nu}],
$
which implies, 

\begin{figure}[t]
\centering
\hspace*{-4mm}
\epsfysize=7 cm \epsfbox{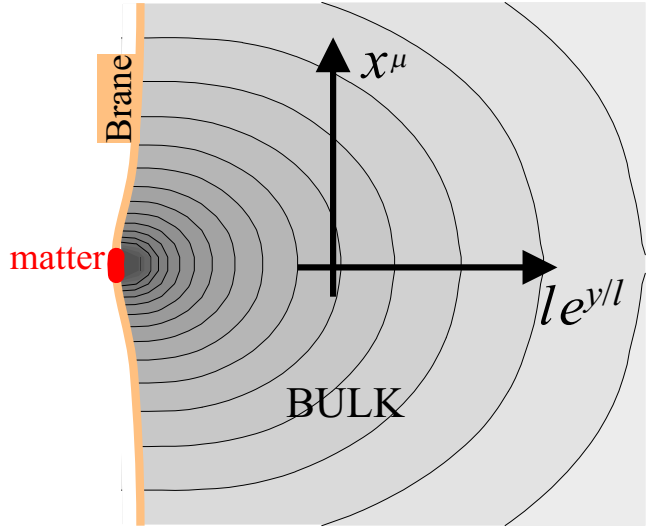}\\[3mm]
\label{fig1}
\narrowtext
\caption[fig1]{Gravitational field of a 
spherically symmetric static source in the Randall-Sundrum gauge.}
\end{figure}
\vspace*{-5mm}
\vspace*{5mm}

\begin{equation}
(\partial_y+2\ell^{-1}) \bar h_{\mu\nu}=
-\kappa\left(T_{\mu\nu}-{1\over 3}\gamma_{\mu\nu}T\right). \quad\quad 
  (\bar y=0+)
\label{jnct}
\end{equation}
Here $\gamma_{\mu\nu}=e^{-2|y|/\ell}\eta_{\mu\nu}$ is the background
spatial metric.
The condition (\ref{jnct}) can now be expressed in the RS gauge.
Since $h_{55}$ and $h_{5\mu}$ vanish in both gauges, 
the most general transformation between them must take
the form
\begin{eqnarray}
&&\xi^5=\hat \xi^5(x^{\rho}),\cr
&&\xi^{\mu}={-\ell\over 2}
\gamma^{\mu\nu}\hat \xi^5(x^{\rho})_{,\nu}+
\hat \xi^{\mu}(x^{\rho}),
\label{synchronous}
\end{eqnarray}
where $\hat\xi^5$ and $\hat\xi^{\mu}$ 
are independent of $y$, and for the moment we are not assuming 
the condition (\ref{ywall}). 
From the gauge 
transformation equations
\begin{equation}
h_{\mu\nu} = \bar h_{\mu\nu} - \ell \hat \xi^5{}_{,\mu\nu}
 -2 \ell^{-1}\gamma_{\mu\nu}\hat \xi^5 + \gamma_{\rho(\mu}
 \hat \xi^{\rho}{}_{,\nu)}, 
\label{gt}
\end{equation}
the junction condition (\ref{jnct}) becomes
\begin{equation}
(\partial_y+2\ell^{-1}) h_{\mu\nu} = -\kappa\Sigma_{\mu\nu},
\quad\quad (y=0+) \label{jump2}
\end{equation}
where we have introduced the combination
\begin{equation}
\Sigma_{\mu\nu}=
\left(
{T_{\mu\nu}-{1\over 3} \gamma_{\mu\nu} T} 
\right) 
+2 \kappa^{-1}\hat \xi^5_{,\mu\nu}.\label{source}
\end{equation}
This combination, which in some sense includes
the ``bending'' of the wall $\hat \xi^5$, will play the role of the
source term in the RS gauge.

Our solutions must be even under parity, and so from (\ref{jump2})
the derivative of the metric perturbation 
will be discontinuous at the wall.
Combining (\ref{jump2}) with (\ref{rseq}), the equations of motion
in the RS gauge become
\begin{equation}
\left[a^{-2}\Box^{(4)} + \partial_y^2 -4\ell^{-2}+ 4\ell^{-1}
\delta(y)\right]
h_{\mu\nu}=-2\kappa \Sigma_{\mu\nu}\delta(y),
\label{eom}
\end{equation}
where the delta function terms will enforce the discontinuities.
Of course, in order to solve (\ref{eom}),
we must first determine the function $\hat\xi^5$, which enters in the 
definition of the source term $\Sigma$. This function is given
by (\ref{ywall}), as we shall now explain.
Let us define the 5D retarded Green's function, which satisfies 
\begin{eqnarray}
&&\left[a^{-2}\Box^{(4)} + \partial_y^2 -4\ell^{-2}+4\ell^{-1}\delta(y)\right]
G_R(x,x')\cr&&\hspace{3cm}= \delta^{(5)}(x-x').
\label{Gdef}
\end{eqnarray}
The formal solution of (\ref{eom}) is then given by
\begin{equation}
 h_{\mu\nu}(x)=-2\kappa \int d^4 x' G_R(x,x') \Sigma_{\mu\nu}(x'),
\label{formalsol}
\end{equation}
where integration is taken over the $y=0$ surface.
Since $h^{\mu}{}_{\mu}$ must vanish, 
we must impose $\Sigma^{\mu}{}_{\mu}=0$, 
which implies the ``equation of motion'' (\ref{ywall}) for $\hat\xi^5$.
With this choice of $\hat \xi^5$, it is easy to check 
that $h_{\mu\nu}$ given by Eq.(\ref{formalsol}) satisfies
the harmonic condition  $h_{\mu}{}^{\nu}{}_{,\nu}=0$.

The behaviour of $h_{\mu\nu}$ at infinity is
determined by the form of $G_R(x,x')$.
The Green's function can be constructed from a complete set 
of eigenstates in the usual way. Following \cite{RS2}, we have
\begin{equation}
G_R(x,x')=-\int {d^4 k \over (2\pi)^4}
e^{ik_{\mu}(x^{\mu}-x'{}^{\mu})}\Biggl[
{a(y)^2 a(y')^2 \ell^{-1}\over {\bf k}^2-(\omega+i\epsilon)^2}
+\int_0^{\infty} dm\,
 {u_m(y) u_m(y')\over m^2+{\bf k}^2-(\omega+i\epsilon)^2}\Biggr],
\end{equation}
where the first term corresponds to the zero mode and the rest
corresponds to the continuum of KK modes
$ u_m(y)=\sqrt{m\ell/2}$ $\{J_1(m\ell)Y_2(m\ell/a) - 
Y_1(m\ell) J_2(m\ell/a)\}$ $
/\sqrt{J_1(m\ell)^2+Y_1(m\ell)^2}. $
For the stationary case, it is more illustrative to consider the
Green's function for the Laplacian operator, 
which is related to the previous one through
\begin{equation}
G({\bf x},y,{\bf x}',y') = \int_{-\infty}^{\infty} dt' G_R(x,x').
\label{eq12p}
\end{equation}
Here ${\bf x}$ are spatial cartesian coordinates on the wall.
When both points are taken on the wall ($y=y'=0$), we have
\begin{equation}
G({\bf x},0,{\bf x}',0)
\approx {-1\over 4\pi \ell r}\left[1+{\ell^2\over 2 r^2}+...\right],
\end{equation}
where $r=|{\bf x}-{\bf x'}|$. 
Also, when one of the points is on the wall, 
the leading behaviour for large separations in any direction is given by 
\begin{equation}
G({\bf x},y,{\bf x'},0 )\approx 
-{a^3\over 8\pi\ell}{2a^2r^2+3\ell^2\over (a^2r^2+\ell^2)^{3/2}},
\label{falloff}
\end {equation}
This means that the metric perturbation decays
rather steeply towards the AdS horizon at $y\to \infty$, i.e. $a\to 0$
(in fact even the relative metric perturbation
$h_{\mu\nu}/a^2$ falls to zero as we move away from the source).
This behaviour is illustrated in Fig. 1.

Since we are interested in the metric on the wall, 
it is convenient to transform back to Gaussian coordinates. 
From (\ref{gt}), we have
$
 \bar h_{\mu\nu}= h^{(m)}_{\mu\nu}+h^{(\xi)}_{,\mu\nu}+ 
\ell\hat \xi^5{}_{,\mu\nu}
 +2 \ell^{-1}\gamma_{\mu\nu}\hat \xi^5 -
 \hat \xi_{(\mu,\nu)}, 
$
where we decomposed $h_{\mu\nu}$ into the part corresponding 
to the matter fields and the part corresponding to the wall 
displacement,
\begin{eqnarray}
&& h^{(m)}_{\mu\nu}=-2\kappa \int d^4 x' G_R(x,x') 
\left(T_{\mu\nu}-{1\over 3}
  \gamma_{\mu\nu} T\right)(x'),
\label{onethird2}
\\
&& h^{(\xi)}=-4\int d^4 x' G_R(x,x') 
\hat \xi^5(x').
\label{onethird}
\end{eqnarray}
Setting $y=0$ and choosing $\hat \xi_{\mu}$ appropriately, 
we end up with 
the rather simple expression
\begin{equation}
 \bar h_{\mu\nu}= h^{(m)}_{\mu\nu}
 +2 \ell^{-1}\gamma_{\mu\nu}\hat \xi^5,
\label{barhsimple}
\end{equation}
which gives the 
metric perturbation on the wall. 

\vspace{5mm}
\centerline{\it i$)$ Spherical symmetry: }
\vspace{5mm}
As a simple application,  
let us now consider the effect of the KK modes on the
metric. We shall restrict attention to the most interesting case of
a static and spherically symmetric source, as this may be related to 
to the final stage of gravitational collapse.
With our assumptions, the energy momentum tensor can be written as
$$
T_{\mu\nu}=\rho(r) u_{\mu} u_{\nu}.
$$
From (\ref{onethird}) with the aid of (\ref{ywall}), we obtain 
\begin{equation}
h^{(\xi)}={4\over 3}\int^r_0 {dr'\over r'{}^2} 
     \int^{r'}_0 dr'' r''{}^2 V(r''), 
\end{equation}
where $V=(\kappa/2) \int G(x,x')\rho(x')d^3x'.$
Then, from (\ref{formalsol}), we have 
\begin{equation}
h_{00}=-{8\over 3} V(r),\quad h_{rr}=-{8\over 3r^3}\int^r_0 dr' r'{}^2 V(r'). 
\end{equation}
The remaining metric components can be found from the requirement that
$h=0$ plus spherical symmetry. Notice that the fall-off properties of
the metric components at $y\to \infty$ are the same as those for the 
Green's function (\ref{falloff}). Hence the field decays quite
steeply away from the wall. It can be checked that
the perturbation of the square of the Riemann tensor behaves as 
$$
 \delta\left(R_{\mu\nu\rho\sigma}R^{\mu\nu\rho\sigma}\right) 
   \propto a^2, 
$$ 
at large $y$  (uniformly for all values of $r$.) 
For comparison, the same quantity behaves as 
$M^2 a^2 (ar)^{-6}$ in the case of the Schwarzschild
black string \cite{CHR}.

In order to find the metric on the brane, we transform
to Gaussian normal coordinates.
When the point is outside the source, we have
\begin{equation}
V\approx -{\kappa M \over 8\pi\ell r}
 \left(1+{\ell^2\over 2 r^2}\right),\quad \hat\xi^5\approx 
{\kappa M\over 24\pi r}
\end{equation}
where $M =\int d^3x \rho$ is the total mass. 
Using (\ref{barhsimple}), we arrive at the result
\begin{eqnarray}
&& \bar h_{00}={2GM\over r}
\left(1+{2\ell^2\over 3 r^2}\right), 
\bar h_{ij}={2GM\over r}
\left(1+{\ell^2\over 3 r^2}\right)\delta_{ij}.
\label{wf}
\end{eqnarray}
It should also be stressed
that the Newtonian potential $\bar h_{00}/2$, which determines 
the attraction of neighbouring bodies, is not the same as $V$ - which
is just proportional to the Green's function $G(x,0)$. The coefficient
in front of the correction $\ell^2/r^2$, due to the KK modes, is different
in both cases, because $\hat \xi^{5}$ is in some sense four dimensional
and contributes only to the zero mode.

Our solution differs from the weak field limit of
the usual 4 dimensional Schwarzschild solution. This seems to
indicate that gravitational collapse of matter on the wall will
not lead to a Schwarzschild black hole, but to a metric 
which has the asymptotic form of the weak field solution 
(\ref{wf}) \cite{EHM}. 

\vspace{5mm}
\centerline{\it ii$)$ Zero mode truncation: }
\vspace{5mm}
In general, in order to obtain the metric perturbation induced on the 
brane, we must first solve Eq.~(\ref{ywall}) for $\hat\xi^5$, feed the 
solution into Eq. (\ref{eom}) for $h_{\mu\nu}$, and then use the gauge 
transformation (\ref{gt}) to obtain $\bar h_{\mu\nu}$. The expectation is 
that this should reproduce the results of linearized Einstein gravity
with some small corrections. Let us now show that, indeed, in the case
of a single brane 
the zero mode truncation of the five dimensional theory coincides with the 
usual linearized four dimensional gravity. 

If both arguments of the two-point function are on the wall, then
$G_R(x,x')$ is dominated by the zero mode
contribution, 
$
G_R(x,x')\approx {\delta^{(4)}(x^{\mu}-x^{\mu}{}')/ \ell\,\Box^{(4)}}.
$
Substituting in (\ref{onethird2}), we find that the 
induced metric on the wall is given by
\begin{equation}
\bar h_{\mu\nu} = -16 \pi G {1\over \Box^{(4)}}
 \left(T_{\mu\nu}-{1\over 2}\gamma_{\mu\nu} T \right),
\end{equation}
where $G = \ell^{-1} G_5$ is the four dimensional Newton's constant.
Thus, we recover the linearized Einstein's equations.

It should be noted, however, that a happy cancellation has occurred: the
factor 1/3 in Eq.(\ref{onethird2}) has turned into the familiar 1/2 
in the process of going to Gaussian normal coordinates (that is,
absorbing $\hat\xi^5$) 
through Eq. (\ref{barhsimple}). 
As we shall see, this cancellation does not occur in the case when we
have two branes, which leads of course to some interesting consequences.

\vspace{5mm}
\centerline{\it iii$)$ Two branes and light deflection: }
\vspace{5mm}
In the case when we have two branes, one with positive tension
at $y=0$ and a second one at $y=d$ with negative tension, 
the previous arguments can
be repeated without any basic formal alterations. The only
differences are that the normalization of the zero mode changes by a 
factor of $(1-e^{-2d/\ell})$ and, more important, the effect of
the ``goldstone'' mode $\hat \xi^5$ does not cancel out.
Following the steps of the derivation given above, we find that
in the zero mode approximation the gravitational field on each 
of the branes satisfies \cite{valery}
\begin{equation}
\left({1\over a^2}\Box^{(4)}\bar h_{\mu\nu}\right)^{(\pm)}
= -\sum_{\sigma=\pm} 16\pi G^{(\sigma)} \left(T_{\mu\nu}-{1\over 3}
  \gamma_{\mu\nu} T\right)^{(\sigma)} 
\pm {16\pi G^{(\pm)}\over 3}{\sinh(d/\ell)\over e^{\pm d/\ell}} 
\gamma_{\mu\nu} T^{(\pm)},
\label{big}
\end{equation}
where the plus and minus refer to quantities on the wall with positive
and with negative tension respectively. 
Here, we have introduced 
\begin{equation}
G^{(\pm)}={G_5\ell^{-1} e^{\pm d/\ell}\over 2 \sinh(d/l)}
\end{equation}
which plays the role of Newton's constant in a Brans-Dicke (BD)
parametrization (we follow the 
conventions of Ref. \cite{will}).

Strictly speaking, this parametrization holds when ``the other'' wall
(the one in which we do not live) is empty. Let us first consider this
situation. In this case, the BD parameter is given by
\begin{equation}
\omega_{BD}^{(\pm)}= {3\over 2} (e^{\pm 2d/l} -1).
\end{equation}
Observations require that $\omega_{BD}>3000$ \cite{will}.
In the positive tension brane, this is achieved with $d/l > 4$,
and we have an acceptable gravity theory even without 
stabilizing the dilaton. In the negative tension brane, we
find that the BD parameter is always negative but greater than
-3/2. In the Einstein frame, the kinetic term for the 
BD field has the usual sign for $\omega_{BD}>-3/2$. This 
suggests that the system of two branes is well behaved in spite 
of the negative tension in one of the branes. 

Now, let us consider the effect of ``shadow'' matter, which lives on 
the other membrane. This appears only in the
first term in (\ref{big}). Hence, for non-relativistic matter,
and assuming spherical symmetry, its contribution to the Newtonian potential 
$\bar h_{00}$ will be twice 
as large as its contribution to any of the diagonal spatial 
components, say $\bar h_{zz}$. 
This is in contrast with the situation in Einstein's theory, where
the contribution to the Newtonian potential is the same as the
contribution to $\bar h_{zz}$. For a source in the $x,y$ plane,
the deflection of a light ray
travelling in the $y$ direction
is given by $\ddot x = (1/2)(\bar h_{00}+\bar h_{yy})_{,x}$. Therefore, for
the same Newtonian mass the deflection of light rays caused by
shadow matter is 25\% smaller that in Einstein gravity.
It would be interesting to investigate this possible effect in an 
astrophysical context. This is left for future research.

\centerline{\bf Acknowledgements}

This work was done while participating in the workshop on 
Structure Formation in the Universe, at the Isaac Newton Institute.
We thank the organizers for their hospitality, and the participants
for numerous conversations. We owe special thanks to M. Bucher, 
T. Chiba, S.W.~Hawking, R.~Gregory, K. Maeda,
X. Montes, A. Pomarol, O. Pujolas, 
V.A.~Rubakov, H. Reall, M. Sasaki, T. Shiromizu
and N. Turok. J.G. acknowledges support from CICYT, 
under grant AEN98-1093. T.T. acknowledges support
from Monbusho System to Send Japanese Researchers Overseas.


\begin{thebibliography}{99}
\bibitem{effort}
The RS gauge can be chosen in the bulk for almost the same reason 
that allows the choice of the synchronous tranverse and traceless gauge 
in the cosmological context (in the absence of matter sources). 
In fact, there is some residual gauge freedom generated by 
(\ref{synchronous}) with $\hat \xi^{\nu}{}_{,\nu}{}^{,\mu}+2
\hat \xi^{\mu}{}_{,\nu}{}^{,\nu}=0$ and
$\hat \xi^5=\ell\hat\xi^{\mu}{}_{,\mu}/4$.
\bibitem{valery} In an earlier version of this paper
the sign of the last term of Eq. (\ref{big}) in the case of a negative 
tension brane was not correct. This was pointed out to us by V.A. Rubakov.
\end{thebibliography}
\end{document}